\documentclass[12pt]{article}
\usepackage{epsfig}

\newcommand{\tr}{{\rm tr}}
\newcommand{\pfrac}[2]{\left(\frac{#1}{#2}\right)}

\begin{document}
\thispagestyle{empty}
\begin{flushright}
MZ-TH/01-06\\
hep-ph/0103047\\
March 2001\\
\end{flushright}
\vspace{0.5cm}
\begin{center}
{\Large\bf Low-energy gluon contributions to the\\[7pt]
vacuum polarization of heavy quarks}\\
\vspace{16pt}
{\large S.~Groote$^1$ and A.A.~Pivovarov$^{1,2}$}\\[12pt]
$^1$ Institut f\"ur Physik der Johannes-Gutenberg-Universit\"at,\\[3pt]
Staudinger Weg 7, 55099 Mainz, Germany\\[7pt]
$^2$ Institute for Nuclear Research of the\\[3pt]
  Russian Academy of Sciences, Moscow 117312
\end{center}
\vspace{1cm}
\begin{abstract}\noindent
We calculate a correction to the electromagnetic current induced by a heavy
quark loop. The contribution of this correction to the vacuum polarization
function appears at the $O(\alpha_s^3)$ order of perturbation theory and has
a qualitatively new feature -- its absorptive part starts at zero energy in
contrast to other contributions where the absorptive parts start at the
two-particle threshold. Our result imposes a constraint on the order $n$ of
the moments used in the heavy-quark sum rules, $n<4$.
\end{abstract}

\vspace{3pt}

\vspace{3pt}

\newpage

High precision tests of the standard model remain one of the main topics of
particle phenomenology. The recent observation of a possible signal from the
Higgs may complete the experimentally confirmed list of the standard model
particles~\cite{Higgs}. Because experimental data are becoming more and more
accurate, the determination of numerical values of the parameters of the
standard model lagrangian will require more accurate theoretical formulae.
Recently an essential development in high-order perturbation theory
calculations has been observed. A remarkable progress has been made in the
heavy quark physics where a number of new physical effects have been
described theoretically with high precision. The cross section of top--antitop
production near the threshold has been calculated at the
next-to-next-to-leading order of an expansion in the strong coupling constant
and velocity of a heavy quark with an exact account for Coulomb interaction
(as a review, see Ref.~\cite{Hoang:2000yr}). This allows for the best
determination of a numerical value of the top quark mass. The method of
Coulomb resummation resides on a nonrelativistic approximation for the Green
function of the quark-antiquark system near the threshold and has been
successfully used for the heavy quark mass determination within the sum
rules~\cite{vol,leut,volnew}. Being applied to quarkonium systems this method
is considered to give the best estimates of heavy quark mass
parameters~\cite{PeninX,melye,hoamom,sumino}. Technically an enhancement of
near-threshold contributions to sum rules is achieved by considering integrals
of the spectral density of the heavy quark production with weight functions
which suppress the high-energy tail of the spectrum. The integrals with weight
functions $1/s^n$ for different positive integer $n$, $s=E^2$, where $E$ is
the total energy of the quark-antiquark system, are called moments of the
spectral density and most often used in the sum rules analysis~\cite{largen}. 

In the present note we show that there is a strong constraint on the order $n$
of the moment that can be used in heavy quark sum rules. Because of the
contribution of low-energy gluons, only moments with $n<4$ exist if
theoretical expressions for the corrrelators include the $O(\alpha_s^3)$ order
of perturbation theory.

The basic quantity for the analysis within sum rules is a vacuum polarization
function $\Pi(q^2)$
\begin{equation}\label{first}
12\pi^2i\int\langle Tj_\mu(x)j_\nu(0)\rangle e^{iqx}d^4x
  =(q_\mu q_\nu-g_{\mu\nu}q^2)\Pi(q^2)
\end{equation}
of the vector current $j^\mu=\bar q\gamma^\mu q$ of a heavy fermion $q$. With
the spectral density $\rho(s)$ defined by the relation 
\begin{equation}\label{firspe}
\rho(s)=\frac1{2\pi i}(\Pi(s+i0)-\Pi(s-i0)),\quad s>0
\end{equation}
the dispersion representation
\begin{equation}\label{disprel}
\Pi(q^2)=\int\frac{\rho(s)ds}{s-q^2}
\end{equation}
holds. A necessary regularization (subtractions, for instance) is assumed in
Eq.~(\ref{disprel}). The normalization of the vacuum polarization function 
$\Pi(q^2)$ in Eq.~(\ref{first}) is chosen such that one obtains a high-energy
limit $\lim_{s\to\infty}\rho(s)=1$ for a lepton. For the quark in the
fundamental representation of the $SU(N_c)$ gauge group a high-energy limit
of the spectral density reads $\rho(\infty)=N_c$. The integral in
Eq.~(\ref{disprel}) runs over the whole spectrum of the correlator in
Eq.~(\ref{first}) or over the whole support of the spectral density $\rho(s)$
in Eq.~(\ref{firspe}). The moments of the spectral density $\rho(s)$ of the
form 
\begin{equation}\label{mom}
{\cal M}_n=\int\frac{\rho(s)ds}{s^{n+1}}
\end{equation}
are usually studied within the sum rules method for heavy quarks~\cite{largen}.
These moments are related to the derivatives of the vacuum polarization
function $\Pi(q^2)$ at the origin,
\begin{equation}\label{momder}
{\cal M}_n=\frac1{n!}\pfrac{d}{dq^2}^n\Pi(q^2)\bigg|_{q^2=0}.
\end{equation}
Such moments are chosen in order to suppress a high energy part of the
spectral density $\rho(s)$ which is not measured accurately in the experiment.
Within the sum rule method one assumes that the moments in Eq.~(\ref{mom}) can
be calculated for any $n$ or, equivalently, that the derivatives in
Eq.~(\ref{momder}) exist for any $n$. The existence of moments seems to be
obvious because one implicitly assumes that the spectral density of the heavy
quark electromagnetic currents $\rho(s)$ vanishes below the two-particle
threshold $s=4m^2$, which means that the vacuum polarization function of
heavy quarks $\Pi(q^2)$ is analytic in the whole complex plane of $q^2$ except
for the cut along the positive real axis starting from $4m^2$. This assumption
about the analytic properties of the vacuum polarization function $\Pi(q^2)$
is known to be wrong if a resummation of Coulomb effects to all orders of
perturbation theory is performed: as a result of such a resummation the
Coulomb bound states appear below the perturbation theory threshold $s=4m^2$.
The assumption that the moments in Eq.~(\ref{mom}) exist for any $n$ may also
be wrong in high orders of perturbation theory in models with massless
particles, for example, in QCD with massless gluons. The validity of this
assumption depends on details of the interaction. In QCD, at the
$O(\alpha_s^3)$ order of perturbation theory there is a contribution of
massless states to the correlator in Eq.~(\ref{first}) that leads to the
infrared (small $s$) divergence of moments for large $n$ because of the
branching point (cut) singularity of $\Pi(q^2)$ at the origin. We determine
the behaviour of the vacuum polarization function $\Pi(q^2)$ at small $q^2$
($q^2\ll m^2$) as 
\begin{equation}\label{res}
\Pi(q^2)|_{q^2\approx 0}=C_g\pfrac{q^2}{4m^2}^4\ln\pfrac{\mu^2}{-q^2}
\end{equation}
with 
\begin{equation}\label{res1}
C_g=\frac{17}{243000}d_{abc}d_{abc}\pfrac{\alpha_s}\pi^3.
\end{equation}
Here $d_{abc}$ are the totally symmetric structure constants of the $SU(N_c)$
gauge group defined by the relation $d_{abc}=2\tr(\{t^a,t^b\}t^c)$, and $t^a$
are generators of the group with normalization $\tr(t^a t^b)=1/2$. For the
$SU(3)$ gauge group of QCD one has $d_{abc}d_{abc}=40/3$. The parameter $\mu$
in Eq.~(\ref{res}) is the renormalization point.

The singularity of the vacuum polarization function given in Eq.~(\ref{res})
(a cut along the positive real axis in a complex $q^2$-plane) prevents one
from calculating moments of the spectral density in Eq.~(\ref{mom}) with
$n>4$. Indeed, the high order derivatives of $\Pi(q^2)$ at the origin
determining the high order moments according to Eq.~(\ref{momder}) do not
exist for $n>4$ because of a branching point singularity as one can see from
Eq.~(\ref{res}). In terms of the moments one can see this by calculating the
behaviour of the spectral density at small energy $s$,
\begin{equation}\label{spect}
\rho(s)|_{s\approx 0}=C_g\pfrac{s}{4m^2}^4
\end{equation}
which makes integrals in Eq.~(\ref{mom}) divergent at small $s$ for $n>4$.
The formulae for the vacuum polarization function in Eqs.~(\ref{res})
and~(\ref{res1}) are given for a heavy quark in the $SU(N_c)\otimes U(1)$
gauge model. The result for QED may be obtained with the obvious substitution
$\alpha_s\to\alpha$ for the coupling constant and by setting
$d_{abc}d_{abc}=1$. Contributions of light (massless) quarks appear in the
$O(\alpha_s^4)$ order of perturbation theory and are neglected. 

We now present a derivation of our result given in Eqs.~(\ref{res})
and~(\ref{res1}) and discuss some consequences for the phenomenology of heavy
quarks. Note that the induced current is a correction of order $1/m^4$ in the
inverse heavy quark mass which vanishes in the limit of an infinitely heavy
quark. Corrections in inverse heavy quark masses are important for tests of
the standard model at the present level of precision and have been already
discussed in various areas of particle
phenomenology~\cite{PivovarovZ,chetm,larinm}.

\begin{figure}\begin{center}
\epsfig{figure=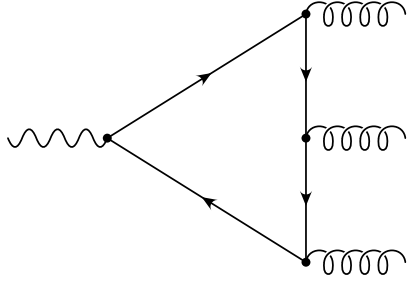, scale=0.4}
\caption{\label{fig1}Heavy quark loop correction to the electromagnetic
  current}\end{center}
\end{figure}

A diagram which gives a correction to the electromagnetic current due to a
heavy quark loop is given in Fig.~\ref{fig1}. Two-gluon transitions are
forbidden according to a generalization of Farry's theorem to nonabelian
theories~\cite{Farri}. We are interested in a behaviour of the amplitude
associated with the diagram Fig.~\ref{fig1} at low energy and therefore take
the limit  of a very heavy quark. Formally the limit $m\to\infty$ is taken
which in physical terms means that $m$ is much larger than all momenta of
external legs of the diagram, namely the three gluons and the photon. The
induced current $J^\mu$ is written in a covariant form as a derivative of an
antisymmetric operator ${\cal O}_{\mu\nu}$ built from the gluon fields only,
\begin{equation}
J_\mu=\partial_\nu{\cal O}_{\mu\nu},\qquad
{\cal O}_{\mu\nu}+{\cal O}_{\nu\mu}=0.
\end{equation}
This structure of the induced current automatically guarantees the current
conservation
\begin{equation}
\partial_\mu J^\mu=0
\end{equation}
as it should be for the electromagnetic current.

A straightforward calculation of the diagram presented in Fig.~\ref{fig1}
gives the result for the induced correction to the electromagnetic current
\begin{equation}\label{fin1}
J^\mu=\frac{-g_s^3}{1440\pi^2 m^4}\left(5\partial_\nu{\cal O}_1^{\mu\nu}
  +14\partial_\nu{\cal O}_2^{\mu\nu}\right)
\end{equation}
with
\begin{equation}\label{fin2}
{\cal O}_1^{\mu\nu}=d_{abc}G^a_{\mu\nu}G^b_{\alpha\beta}G^c_{\alpha\beta},
  \quad
{\cal O}_2^{\mu\nu}=d_{abc}G^a_{\mu\alpha}G^b_{\alpha\beta}G^c_{\beta\nu}
\end{equation}
where $G^a_{\mu\alpha}$ is a gauge field strength tensor for the
gauge group $SU(N_c)$.

A correlator of the induced current $J^\mu$ has a general form
\begin{equation}\label{glucorr}
\langle TJ_\mu(x)J_\nu(0)\rangle=-\partial_\alpha\partial_\beta 
\langle T{\cal O}_{\mu\alpha}(x){\cal O}_{\nu\beta}\rangle
\end{equation}
where an explicit expression of the current as a derivative of the
antisymmetric operator ${\cal O}_{\mu\nu}$ has been employed. The resulting
correlator $\langle T{\cal O}_{\mu\alpha}(x){\cal O}_{\nu\beta}\rangle$ in
Eq.~(\ref{glucorr}) contains only gluonic operators as is seen from
Eqs.~(\ref{fin1}) and~(\ref{fin2}). Such correlators were considered
previously in the framework of perturbation theory~\cite{PivovarovU,Okada}.
In the leading order of perturbation theory the correlator in
Eq.~(\ref{glucorr}) has a topological structure of a sunset diagram.
Technically, a convenient procedure of computing the sunset-type diagrams is
to work in the configuration space~\cite{GrooteV}. We find 
\begin{equation}\label{xspace}
\langle T J_\mu(x)J_\nu(0)\rangle
  =-\frac{34}{2025\pi^4m^8}\pfrac{\alpha_s}\pi^3d_{abc}d_{abc}
  \left(\partial_\mu\partial_\nu-g_{\mu\nu}\partial^2\right)\frac1{x^{12}}
\end{equation}
A Fourier transform of the correlator in Eq.~(\ref{xspace}) gives the vacuum
polarization function in momentum space which reads
\begin{equation}\label{fourier}
12\pi^2i\int\langle TJ_\mu(x)J_\nu(0)\rangle e^{iqx}d^4x
  = C_g(q_\mu q_\nu-g_{\mu\nu}q^2)\pfrac{q^2}{4m^2}^4\ln\pfrac{\mu^2}{-q^2}
\end{equation}
with the constant $C_g$ taken from Eq.~(\ref{res1}). The spectral density of
the polarization function in Eq.~(\ref{fourier}) is given in Eq.~(\ref{spect}).

Note that the spectral density of the correlator in Eq.~(\ref{xspace}) can be
found without an explicit calculation of its Fourier transform. Instead one
can use a spectral decomposition (dispersion 
representation) in configuration space,
\begin{equation}
\frac{i}{x^{12}}=\frac{\pi^2}{2^8\Gamma(6)\Gamma(5)}\int_0^\infty s^4D(x^2,s)ds
\end{equation}
with $D(x^2,s)$ being the propagator of a scalar particle of mass $\sqrt s$,
\begin{equation}
D(x^2,m^2)=\frac{im\sqrt{-x^2}K_1(m\sqrt{-x^2})}{4\pi^2(-x^2)}
\end{equation}
where $K_1(z)$ is a McDonald function (a modified Bessel function of the third
kind, see e.g.\ Ref.~\cite{mac}). $\Gamma(z)$ is Euler's gamma function.

An asymptotic behaviour of the spectral density of the corresponding
contribution for large energy (when the limit of massless quarks can be used)
is well known~\cite{kataev,chet,surg} and reads
\begin{equation}\label{asym}
\pfrac{\alpha_s}\pi^3\frac{d_{abc}d_{abc}}{1024}
  \left(\frac{176}3-128\zeta(3)\right).
\end{equation}
Here $\zeta(z)$ is the Riemann $\zeta$ function and $\zeta(3)=1.20206\ldots$.
The contribution to the spectral density given in Eq.~(\ref{asym}) is negative
while our result given in Eq.~(\ref{spect}) is positive as it should be for
the spectral density of the electromagnetic current which is an Hermitian
operator.

In QCD we find 
\begin{equation}\label{res2}
\Pi_{\rm QCD}(q^2)|_{q^2\approx 0}=\frac{17}{18225}
  \pfrac{\alpha_s}\pi^3\pfrac{q^2}{4m^2}^4\ln\pfrac{\mu^2}{-q^2}.
\end{equation}

Our result has an immediate application to the determination of heavy quark
parameters within the method of sum rules. Because of the low-energy gluon
contributions, the large $n$ ($n>4$) moments of the spectral density do not
exist and cannot be used for phenomenological analyses. Note that in early
considerations of sum rules quite large $n$ were used. For instance, the
numerical value of the gluon condensate was extracted from sum rules for the
moments with $n\sim 10\div 20$~\cite{largen,rad}. In view of our result one
has either to limit  the accuracy of theoretical calculations for the
moments to the $O(\alpha_s^2)$ order of perturbation theory which seems
insufficient for a high precision analysis of quarkonium systems (especially
if the Coulomb resummation in all orders is performed) or to use only a few
first moments with $n<4$. For small $n$, however, the high-energy
contribution, which is not known experimentally with a reasonable precision,
is not sufficiently suppressed and introduces a large quantitative uncertainty
into sum rules for the moments. An alternative analysis based on finite energy
sum rules is free from such a problem and can be used in phenomenological
applications~\cite{KrasnikovZ}.

Note in passing that there is no low-energy gluon contribution (and low-energy
divergence problem) for correlators of the currents containing only one heavy
quark with mass $m$. The spectrum of such correlators starts at $m^2$ and
there are no massless intermediate states contributing to the correlator in
perturbation theory. The theoretical expressions for such correlators can be
used for high precision tests when the accuracy of experimental data in
correspondent channels will improve in the future.

To conclude, we have presented a correction to the electromagnetic current of
a heavy quark induced by a virtual heavy quark loop. The spectrum of the
correlator of such an induced current starts at zero energy. This fact makes
impossible the standard analysis of the moment sum rules for $n>4$ at 
the $O(\alpha_s^3)$ order of perturbation theory.

The work is partially supported by the Russian Fund for Basic Research under
contract 99-01-00091. A.A.~Pivovarov is an Alexander von Humboldt fellow.
S.~Groote acknowledges a grant given by the Deutsche Forschungsgemeinschaft,
Germany.

\end{document}